\documentclass{article}
\usepackage{amssymb}
\usepackage{amsbsy}
\newcommand{\be}{\begin{equation}}
\newcommand{\ee}{\end{equation}}

\def\bea{\begin{eqnarray}}
\def\eea{\end{eqnarray}}
\def\bean{\begin{eqnarray*}}
\def\eean{\end{eqnarray*}}

\newcommand{\barr}{\begin{array}}
\newcommand{\earr}{\end{array}}

\newcommand{\bed}{\begin{displaymath}}
\newcommand{\eed}{\end{displaymath}}
\newcommand{\bal}{\begin{array}{ll}}
\newcommand{\eal}{\end{array}}

\def\bvec#1{\raise1.5ex\hbox{$\rightarrow$}\mkern-16.5mu #1}

\def\m#1{\mathcal#1}

\newcommand{\bs}{\boldsymbol}

\begin{document}

\title{\hfill ~\\[-30mm]
       \hfill\mbox{\small UFIFT-HEP-07-8}\\[30mm]
       \textbf{Tri-Bimaximal Neutrino Mixing and the Family Symmetry
       $\boldsymbol{\m Z_7 \rtimes \m Z_3}$}}  
\date{}
\author{\\Christoph Luhn,\footnote{E-mail: {\tt luhn@phys.ufl.edu}}~~
        Salah Nasri,\footnote{E-mail: {\tt snasri@phys.ufl.edu}}~~
        Pierre Ramond,\footnote{E-mail: {\tt ramond@phys.ufl.edu}}\\ \\
  \emph{\small{}Institute for Fundamental Theory, Department of Physics,}\\
  \emph{\small University of Florida, Gainesville, FL 32611, USA}}

\maketitle

\begin{abstract}
\noindent The Non-Abelian finite group $\m P \m S \m L_2(7)$ is the only {\em
  simple} subgroup of $SU(3)$ with a complex three-dimensional irreducible 
representation. It has two maximal subgroups, $\m S_4$ which,  along with its
own $\m A_4$ subgroup, has been successfully applied in numerous models of
flavor, as well as the $21$ element Frobenius group $\m Z_7 \rtimes \m Z_3$,
which has gained much less attention. We show that it can also be used to
generate tri-bimaximal mixing in the neutrino sector, while allowing for  
quark and charged lepton hierarchies.
\end{abstract}

\thispagestyle{empty}
\vfill
\newpage
\setcounter{page}{1}


\section{Introduction}
The triplication of chiral families found in Nature remains a daunting mystery in spite of numerous data in the form of quark mixings, and quark and charged lepton masses. The last decade has witnessed a  spectacular addition with the discovery of neutrino oscillations. Remarkably, the resulting lepton mixings are quite distinct from those obeyed by the quarks, in spite of the suggestive Pati-Salam quark-lepton unification. The most arresting feature of the neutrino mixing data is the appearance of a small angle, flanked by two large ones. 

The natural expectation from a Grand-Unified Theory such as $SO(10)$ was the other way around: {\em one} large and two small angles. The Froggatt-Nielsen formalism which had proved so promising in the quark sector does not seem to account for the apparent fine-tuning of the prefactors of the matrix elements. 

One way to account for the prefactors is to appeal to a family symmetry. The
large value of the top quark mass suggests an $SU(3)$ (not $SO(3)$) family
group, with the three chiral families belonging to its  triplet
representation. However such a scheme is fraught with anomalies, and
introduces a very complicated and unlikely Higgs structure
\cite{Curtright:1979uk}. 
Recently, spurred on
by the neutrino mixing data, many authors have argued for a {\em finite}
family group 
\cite{Pakvasa:1977in,
Ma:2001dn,
Hagedorn:2006ug,
deMedeirosVarzielas:2005ax,
Ma:2006ip,
Kaplan:1993ej,
Ma:2007ia
}, either a subgroup of $SO(3)$ or of $SU(3)$ \cite{Dickson}. 

Particularly intriguing is the remark that a very good approximation to lepton
mixing is given by the tri-bimaximal mixing matrix
\cite{Harrison:2002er}, 
a pretty matrix with an
ugly name. It is natural to seek theories where, to first
approximation, the lepton mixing matrix (MNSP) is tri-bimaximal, and the quark
mixing matrix (CKM) is unity. Deviations from this would occur when Cabibbo corrections are turned on. Phenomenologically, this picture suggests a Wolfenstein-like expansion of the MNSP matrix about the tri-bimaximal matrix. Several authors have noted that the tri-bimaximal matrix occurs naturally in Non-Abelian finite groups. These groups have not been systematically studied, and it is the purpose of this letter to partially alleviate this state of affairs in pointing out the existence of an important finite simple group which is a subgroup of continuous $SU(3)$. Its study leads not only to well-researched finite groups such as $\m S_4$ and $\m A_4$, the groups of permutations and even permutations on four objects, respectively, but also to the $21$-element Frobenius group, the semi-direct product of the Abelian rotation groups $\m Z_7$ and $\m Z_3$. 

We present a model based on this group which reproduces naturally
tri-bimaximal mixing {\em and} the  normal hierarchy among neutrino masses,
while at this level of approximation there is no quark mixing.  This scheme is
to be viewed as a starting point in an expansion in Cabibbo-size parameters
which yield the CKM matrix among the quarks, and the so-called ``Cabibbo Haze"
among the leptons \cite{Datta:2005ci}.  
We do not dwell on the alignment of the flavor (familon) fields, as it is
similar to that of the already studied $\Delta(27)$
\cite{deMedeirosVarzielas:2005ax}. 
Tri-bimaximal mixing is uniquely fixed as long as the lepton masses are not
degenerate. Yet the Froggatt-Nielsen approach suggests that the two lightest
leptons are massless in the absence of Cabibbo mixing. For us, the deviation
of the CKM matrix from unity is not obviously correlated with the masses, thus
making the phenomenologically successful Gatto, Sartori, Tonin, and Oakes
\cite{Gatto:1968ss} 
relation between the Cabibbo angle and the down and strange quark mass ratio difficult to explain.


\section{Tri-Bi Maximal Mixing}
Ever since Harrison, Perkins and Scott \cite{Harrison:2002er} 
suggested that the neutrino mixing data could be approximated in terms of the tri-bimaximal matrix

$$\m U^{}_{MNSP}~\approx~\m U^{}_{\m{TB}}~=~\pmatrix {\sqrt{\frac{2}{3}}&\frac{1}{\sqrt{3}}&0\cr -\frac{1}{\sqrt{6}}&\frac{1}{\sqrt{3}}&-\frac{1}{\sqrt{2}}\cr 
-\frac{1}{\sqrt{6}}&\frac{1}{\sqrt{3}}&\frac{1}{\sqrt{2}}}\ ,$$
there have been many proposals to explain its structure in terms of a discrete flavor symmetry. Neglecting phases, the general symmetric matrix that is diagonalized by $\m
U^{}_{\m{TB}}$ can be written in terms of three parameters $r,s,t$

$$
\m M_{\m{TB}}^{} ~=~ 
\pmatrix{ 
r  & s & s \cr 
s  & t & r+s-t  \cr
s  & r+s-t & t}= \m U^{}_{\m{TB}}\pmatrix{r-s&0&0\cr 0&r+2s&0\cr 0&0&2t-r-s}\m U^{T}_{\m{TB}}\ ,$$
where the superscript $T$ means transposition.

We see that there are three criteria for a symmetric matrix $\m M$ to be uniquely diagonalizable by tri-bimaximal mixing, namely

$$\m M^{}_{12}=\m M^{}_{13}\ ,\quad \m M^{}_{22}=\m M^{}_{33}\ ,\quad \m M^{}_{23}= 
\m M^{}_{11}+\m M^{}_{12}-\m M^{}_{22}\ ,$$
as long as it has no degenerate eigenvalues, that is 

$$\m M^{}_{12}\ne 0\ ,\quad \m M^{}_{11}\ne \m M^{}_{33}\ ,\quad 3\m M^{}_{12}\ne  
2\m M^{}_{33}-2\m M^{}_{11}\ .$$

 The down quark mass matrix and that of the charged leptons are closely related in  Grand-Unified Theories. Furthermore, if the
down quark mass matrix is family symmetric,  the mixing coming from the diagonalization of the charged leptons is structurally the same as that of the down quarks, and expected to be small. 
In that case, with the three neutrino masses  given by

$$
m^{}_1~=~r-s\ ,\qquad m_2^{}~=~r+2s\ ,\qquad m^{}_3~=~2t-r-s\ ,$$ we can write the neutrino mass matrix in the form

$$\m M^{}_\nu~\approx~\m M_{\m{TB}}^{}~=~m^{}_1\varphi^{}_{1}\varphi^T_{1}+m^{}_2\varphi^{}_{2}\varphi^T_{2}+m^{}_3\varphi^{}_{3}\varphi^T_{3}\ ,$$
where

$$\varphi^{}_1~=~\frac{1}{\sqrt{6}}\pmatrix{\hfill 2\cr -1\cr -1}\ ,\qquad \varphi^{}_2~=~\frac{1}{\sqrt{3}}\pmatrix{1\cr 1\cr 1}\ ,\qquad\varphi^{}_3~=~\frac{1}{\sqrt{2}}\pmatrix{\hfill 0\cr \hfill 1\cr -1}\ .$$
If all three $m_i$ have similar orders of magnitudes, the tri-bimaximal hypothesis suggests the existence of three familon fields $\varphi_i$ with vacuum values aligned along these three eigenstates. 

Any finite group which reproduces these vacuum alignments is a candidate for explaining the flavor structure of the three chiral families. The quest for models has centered on those discrete groups which reproduce these flavor alignments. 

Remarkably, $\m U_{\m{TB}}$ is ubiquitous among finite groups. Indeed \cite{Lomont}, the tri-bimaximal matrix  is to be found in the smallest non-Abelian discrete group, $\m D_3$ which is the symmetry group of the equilateral triangle. It is also $\m S_3$, the group of permutations on three objects generated by the matrices

$$A=\pmatrix{0&1&0\cr 0&0&1\cr 1&0&0}\ ,\qquad B=\pmatrix{0&0&1\cr 0&1&0\cr 1&0&0}\ .$$
Any $(3\times 3)$ matrix  $S$ which satisfies 

$$S=ASA^{-1}_{}\ ,\qquad S~=~BSB^{-1}_{}\ ,$$
is of the form 

$$S~=~\pmatrix{\alpha&\beta&\beta\cr \beta&\alpha&\beta\cr \beta&\beta&\alpha}\ ,$$ 
and is diagonalized by $\m U_{\m{TB}}$, with two degenerate eigenvalues, corresponding to the reducibility of the the three-dimensional space into  ${\bf 1}+{\bf 2}$, the sum of the $\m D_3$ irreducible representations.

\section{Finite Groups}
We have just seen that, from the neutrino mixing patterns,  there are good reasons to think that a Non-Abelian finite family group lurks behind
the flavor structure of the Standard Model. Unlike continuous groups which have been systematically studied, there is no equivalent body of work for
finite groups, except in the mathematical literature. Mathematicians organize finite groups in two distinct categories, {\em simple} groups, the equivalent
of the simple Lie algebras which generate the continuous Lie groups,  which cannot be decomposed any further, and the rest which can be understood as
conglomerations of  finite simple groups.   All finite groups are composed of finite simple groups. There are two infinite families of finite
simple groups,  groups of Lie type which are generated by Lie group elements over finite fields, and groups of even permutations on five or more
objects. In addition, there is a third (finite) family, namely that of the twenty six {\em sporadic} groups, the largest of which is the ``Monster" with over $10^{54}$ elements,  and the smallest, the Mathieu group $M_{12}$ has $95,040$ elements.

Thankfully, there are only three chiral families in Nature, and  the hunt for candidate finite flavor groups
is limited to those groups which have two- or three-dimensional irreducible
representations. They are to be found among the finite subgroups of $SU(2)$,
the only Lie group with  two-(and three-)dimensional irreducible
representations, and of $SO(3)\approx SU(2)/ \m Z_2$ with real
three-dimensional irreducible representations, and of $SU(3)$, the only Lie
group with a complex three-dimensional irreducible representation. The
classification of the finite subgroups of $SU(3)$ can be found in
\cite{Dickson}. 
A complete list of the finite groups
up to order~32 is given in \cite{Thomas}. 

There are no finite simple groups with two-dimensional irreducible representations, and  {\em only two} finite simple groups
with three-dimensional representations. The smallest is  $\m A_5$, the group of even permutations on five letters, with $60$ elements. It is the symmetry
group of a Platonic solid, the dodecahedron (``Bucky Ball"), and therefore a
finite subgroup of $SO(3)$. Its three-dimensional irreducible representations
are real. The second is $\m P \m S \m L_2(7)$, the projective special linear group of $(2\times 2)$ matrices over $\mathbb F_7$, the finite Galois field of seven elements.  It contains $168$ elements, and has a complex three-dimensional irreducible representation and its conjugate. It is isomorphic to $GL_3(2)$, the group of non-singular $(3\times 3)$ matrices with entries in $\mathbb F_2$. 

$\m P \m S \m L_2(7)$ has six irreducible representations, the singlet
$\bf 1$, the complex $\bf 3$ and its conjugate $\bf\overline 3$, and three real irreducible representations, $\bf 6$, $\bf 7$, and $\bf 8$. Its structure 
closely parallels that of $SU(3)$, except for the reality of the $\bf 6$, and the existence of the $\bf 7$. These representations fit in standard $SU(3)$ representations as shown below (for more details we refer the reader to
\cite{PSLpaper}): 

\vskip .4cm
\begin{center}
{{\begin{tabular}{|c|}
 \hline  \\[-1mm]
~$\bs {SU(3) ~ \supset~\m P \m S \m L_2(7)}$  \hfill\\[-1mm]
   \\
\hline    
 \\[-1mm]
 $(10):~{\bf 3}~=~{\bf 3}\hfill$ \\ 
 $(01):~{\bf \overline 3}~=~{\bf  \overline 3}\hfill$ \\
 $(20):~{\bf 6}~=~{\bf 6}\hfill$ \\
 $(02):~{\bf \overline 6}~=~{\bf 6}\hfill$ \\
$(11):~{\bf 8}~=~{\bf 8}\hfill$ \\ 
$ (30):~{\bf 10}~=~{\bf \overline 3}+{\bf 7}\hfill$ \\
$ (21):~{\bf 15}~=~{\bf 7}+{\bf 8}\hfill$ \\[-1mm]
\\ \hline
\end{tabular}}}\end{center}
\vskip .4cm

\noindent  $\m P \m S \m L_2(7)$ has two maximal subgroups, $\m S_4$ the group of permutations on four objects with $4!=24$ elements, and the semi-direct product group $\m Z_7\rtimes \m Z_3$,  with $21$ elements. 

$$SU(3)\supset {\m P\m S\m L}_2(7)\supset \cases{\m S_4\supset \m A_4\cr \cr \m Z_7\rtimes \m Z_3}$$
The first branch leads to well-studied finite groups, and we refer the reader
to the literature for their detailed properties
\cite{Ma:2001dn,
Hagedorn:2006ug
}. The second branch yields the
so-called Frobenius group which has not been as well studied.
For completeness, we state the embeddings of the corresponding
representations into the representations of ${\m P\m S\m L}_2(7)$
\cite{Hagedorn:2006ug,PSLpaper}: 

\vskip .4cm
\begin{center}
\begin{tabular}{ccc}
\begin{tabular}{|c|}
 \hline  \\[-1mm]
~~${\bf{\bs{\m P \m S \m L_2(7)\supset \m S_4}}}$  \hfill\\[-1mm]
   \\
\hline    
 \\[-1mm]
 ${\bf 3}~=~{\bf 3_2}\hfill$ \\ 
 ${\bf \overline 3}~=~{\bf  3_2}\hfill$ \\
 ${\bf 6}~=~{\bf 1}+{\bf 2}+{\bf  3_1}\hfill$ \\
${\bf 7}~=~{\bf 1'}+{\bf 3_1}+{\bf  3_2}\hfill$ \\ 
$ {\bf 8}~=~{\bf 2}+{\bf 3_1}+{\bf 3_2}\hfill$ \\[-1mm]
\\ \hline
\end{tabular}
~~~~~~~~~~~~~~~
\begin{tabular}{|c|}
 \hline  \\[-1mm]
~~${\bf{\bs{\m P \m S \m L_2(7)\supset \m Z_7\rtimes \m Z_3}}}$ \hfill\\[-1mm]
   \\
\hline    
 \\[-1mm]
 ${\bf 3}~=~{\bf 3}\hfill$ \\ 
 ${\bf \overline 3}~=~{\bf \overline 3}\hfill$ \\
 ${\bf 6}~=~{\bf 3}+{\bf \overline 3}\hfill$ \\
${\bf 7}~=~{\bf 1}+{\bf 3}+{\bf \overline 3}\hfill$ \\ 
$ {\bf 8}~=~{\bf 1'}+{\bf \overline 1'}+{\bf 3}+{\bf\overline 3}\hfill$ \\[-1mm]
\\ \hline
\end{tabular}
\end{tabular}
\end{center}
\vskip .4cm

\subsection{The Frobenius Group $\bs{\m Z_7\rtimes \m Z_3}$}
As the above table implies, this group has five irreducible representations,
the singlet $\bf 1$, two 
conjugate one-dimensional representations the complex $\bf 1'$ and its conjugate $\bf\overline 1'$, as
well as one complex three-dimensional representation $\bf 3$ and its conjugate
$\bf\overline 3$. Their Kronecker products are summarized in the following
table \cite{PSLpaper}:

\vskip .4cm
\begin{center}
{{\begin{tabular}{|c|}
 \hline  \\[-1mm]
~~{\bf$\bs{\m Z_7\rtimes \m Z_3}$  Kronecker Products}\hfill\\[-1mm]
   \\
\hline   
 \\[-1mm]
 ${\bf 1'}\otimes {\bf 1'}~=~{\bf \overline 1'}\hfill$ \\ ${\bf 1'}\otimes {\bf \overline1'}~=~{\bf 1}\hfill$ \\
${\bf 3}\,\,\otimes{\bf 1'}~=~{\bf 3}\hfill$ \\ 
$ {\bf 3}\,\,\otimes{\bf \overline 1'}~=~{\bf 3}\hfill$ \\
 ${\bf 3}\,\,\otimes{\bf 3}\;~=~({\bf 3}+{\bf\overline 3})^{}_s+{\bf \overline 3}_a^{}\hfill$ \\
 ${\bf 3}\,\,\otimes{\bf \overline 3}\;~=~{\bf 1}+{\bf 1'}+{\bf \overline 1'}+{\bf 3}+{\bf \overline3}\hfill$ \\[-1mm]
\\ \hline
\end{tabular}}}\end{center}
\vskip .2cm

\noindent When discussing the vacuum alignment for $\m Z_7\rtimes \m
Z_3$, we will exploit its similarity to the group $\Delta(27)$, which also has
a complex representation $\bf 3 $ and its conjugate $\bf \overline 3$, as well
as nine one-dimensional irreducible representations, and for which the product
${\bf 3}\otimes {\bf 3}$ decomposes into three antitriplets, while ${\bf
  3}\otimes {\bf \overline 3}$ yields the sum of all nine one-dimensional
irreducible representations \cite{Ma:2006ip}. 

It is straightforward to work out the independent invariants built out of two, three and four triplets $\xi=(x,y,z)$ and/or antitriplets
$\overline{\xi}=(\overline{x},\overline{y},\overline{z})$ of $\m Z_7\rtimes \m Z_3$. 
We first construct the invariants obtained from ${\bf 3}\otimes {\bf 3}'\otimes {\bf 3}''$.  The Kronecker
product shows two ways of building an antitriplet from 
${\bf  3}\otimes {\bf 3}'$, the symmetric and the antisymmetric combinations

$$
{\bf \overline 3}_s ~ = ~ \frac{1}{\sqrt{2}} \pmatrix{
z \, y' + y \, z' \cr x \, z' + z \, x' \cr y \, x' + x \, y'} \ , \qquad
{\bf \overline 3}_a ~ = ~ \frac{1}{\sqrt{2}} \pmatrix{
z \, y' - y \, z' \cr x \, z' - z \, x' \cr y \, x' - x \, y'} \ .
$$
Multiplication with the third triplet ${\bf 3}''$ results in two different invariants 

\bean
\frac{1}{\sqrt{2}} \Big( z \, y' x'' + \, x \, z' y'' + \, y \, x' z'' + \,
y \, z' x'' + \, z \, x' y'' + \, x \, y' z''  \Big) \ , \\
\frac{1}{\sqrt{2}} \Big( z \, y' x'' + \, x \, z' y'' + \, y \, x' z'' - \,
y \, z' x'' - \, z \, x' y'' - \, x \, y' z''  \Big) \ ,
\eean
each of which consisting of six terms. Adding and subtracting the two
expressions yields an equally suitable pair of invariants which however
comprises only three terms. Neglecting the overall factor, we have

$$
 z \, y' x'' + \, x \, z' y'' + \, y \, x' z''  \ , \qquad \mathrm{and} \qquad
 y \, z' x'' + \, z \, x' y'' + \, x \, y' z''  \ .
$$
This is the most convenient ``basis'' for the invariants of ${\bf 3}\otimes
{\bf 3}'\otimes {\bf 3}''$ as it has a minimal number of terms. Note also that
both invariants are related to each other by distributing the primes differently,
e.g. changing the roles of $\xi$ and $\xi'$ in the first invariant
gives the second invariant. It is therefore sufficient to just list the
fundamental invariants which generate all other invariants by reordering the
primes. 

In this way, we similarly find the $\m Z_7 \rtimes \m Z_3$ invariants
for the other cubic and quartic products of (anti)triplets

$$
\barr{ll}
{\bf 3} \otimes \overline{\bf 3}:   & 
I^{(2)} ~ = ~ x \, \overline{x} + y \, \overline{y} + z \, \overline{z} \
,\\[2mm]  
{\bf 3} \otimes {\bf 3'} \otimes {\bf 3''}:   &
I^{(3)}_1 ~ = ~ x \, y' z'' + y  \,z' x'' +  z \, x' y'' \ ,\\[1mm] 
{\bf 3} \otimes {\bf 3'} \otimes \overline{\bf 3}:   &
I^{(3)}_2 ~ = ~ x \, x' \overline{y} + y \, y' \overline{z} + z  \, z'
\overline{x} \ ,\\[2mm] 
{\bf 3} \otimes {\bf 3'} \otimes {\bf 3''} \otimes {\bf 3'''}:   &
I^{(4)}_1 ~ = ~ x  \, x' x'' z''' +  y  \, y' y'' x'''+  z  \, z' z'' y''' \
,\\[1mm] 
{\bf 3} \otimes {\bf 3'} \otimes {\bf 3''} \otimes \overline{\bf 3}:   &
I^{(4)}_2 ~ = ~ x \, x' y'' \overline{z} + y \, y' z'' \overline{x} + z  \, z'
x'' \overline{y} \ ,\\[1mm]  
{\bf 3} \otimes {\bf 3'} \otimes \overline{\bf 3} \otimes \overline{\bf3'}:  &
I^{(4)}_3 ~ = ~ x \, x' \overline{x} \, \overline{x}' + y  \,y'\overline{y} \,
\overline{y}' + z \, z' \overline{z} \, \overline{z}' \ ,\\[1mm] 
{\bf 3}\otimes{\bf 3'}\otimes \overline{\bf 3} \otimes \overline{\bf 3}':   &
I^{(4)}_4 ~ = ~ x \, z' \overline{x}  \, \overline{z}' + y \,  x'\overline{y}
\, \overline{x}' + z \,  y' \overline{z} \, \overline{y}' \ . 
\earr
$$

\noindent Shown here is only one possible way of distributing the
primes among different fields on the right-hand side. Taking this multiplicity
of invariants into account,  there exist two invariants of type $I^{(3)}_1$, four of type\footnote{As an
aside, we note that $I^{(4)}_1$ includes Klein's quartic curve, an object mathematicians have been studying for a long time. It
corresponds to the two-dimensional Riemann surface of genus 3 which has the
maximum number of symmetries allowed by its genus.} $I^{(4)}_1$, three of type 
$I^{(4)}_2$ and four of type $I^{(4)}_4$. Unless two or more of the (anti)triplets coincide, all of these invariants are independent and can
therefore enter the Lagrangian with different coefficients. Note however that the {\it square} of the quadratic invariant $I^{(2)}$ can be expressed in
terms of the quartic invariants $I^{(4)}_3$ and $I^{(4)}_4$.

In general, the structure of the invariants of a group affects two sectors:
($i$)~the potential of the familon fields which in turn gives rise to a
certain vacuum alignment, and ($ii$)~the coupling of the Standard Model
fermions to the familon fields which then determines the structure of the mass
matrices.  

\subsection{Vacuum Alignment}
It has been pointed out in
\cite{deMedeirosVarzielas:2005ax} 
that the existence of the
invariant~$I^{(4)}_3$ can readily explain a very powerful vacuum structure of
triplet and antitriplet familon fields~$\varphi$ and $\overline{\varphi}$. 
Allowing for the invariants $I^{(2)}$, $I^{(4)}_3$ and $I^{(4)}_4$ while
forbidding the rest (e.g. with an additional $U(1)$ symmetry), the most general
potential for only one field~$\overline\varphi$ takes the form

$$
\mu^2 \cdot \sum_i \overline{\varphi}^\dagger_i \overline{\varphi}_i  
~+~ \lambda \left( \sum_i \overline{\varphi}^\dagger_i \overline{\varphi}_i
\right)^2  
~+~ \kappa \cdot\sum_i \overline{\varphi}^\dagger_i\overline{\varphi}_i
\overline{\varphi}^\dagger_i\overline{\varphi}_i 
\ . 
$$
Here we have used the fact that the square of $I^{(2)}$ already includes
$I^{(4)}_4$. Depending on the  sign of $\kappa$, we obtain the vacuum configurations

$$
\kappa > 0: ~~ \langle \overline{\varphi} \rangle ~\propto ~
\frac{1}{\sqrt{3}} \pmatrix{1 \cr 1  \cr 1} \ , \qquad
\kappa < 0: ~~ \langle \overline{\varphi} \rangle ~\propto ~ \pmatrix{0 \cr 0 \cr 1} \ .
$$
Compared with the group $\Delta(27)$
applied in
\cite{deMedeirosVarzielas:2005ax}, 
our order~21 group is more minimal with respect
to the number of invariants derived from the product ${\bf 3} \otimes {\bf 3}
\otimes {\bf \overline 3}\otimes {\bf \overline 3}$. While we have only $I^{(4)}_3$ and $I^{(4)}_4$, there is a third independent
type of invariant for the group $\Delta(27)$; this additional invariant   
$x \, x' \overline{y}\, \overline{z}' + y\, y' \overline{z}\, \overline{x}' + z \,z' \overline{x} \,\overline{y}'$ is neither necessary for the
vacuum alignment nor is it mentioned in \cite{deMedeirosVarzielas:2005ax}. 

Since the relevant invariants of the groups $\m Z_7 \rtimes \m Z_3$ and
$\Delta(27)$ are identical, one could construct a complete model of flavor
along the same lines as
\cite{deMedeirosVarzielas:2005ax}. 
This would include coupling
different antitriplet familon fields to each other so that one can generate
the vacuum structure

$$
\langle \overline \varphi \rangle ~\propto~ \frac{1}{\sqrt{2}}  
\pmatrix{\hfill 0 \cr\hfill  1 \cr -1} \ , 
$$
which is one of the three eigenstates of $\mathcal{M_{\mathcal{TB}}}$. Then
one would have to arrange products of $I^{(2)}$ invariants in a sophisticated
manner so that, in the end, all masses and mixing angles of the Standard Model fermions including the
neutrinos are reproduced. Below, we  present an alternative way of building
the Yukawa couplings, in which, to first approximation, only the two
vacuum alignment vectors directly induced by the invariant $I^{(4)}_3$ are
needed. That is, we do not couple different familon fields to each other in the
potential to obtain new alignment vectors.

\section{Mass Matrices}
 Our construction of the Yukawa couplings is predicated on three
 assumptions. One is that the Higgs fields which break the gauge symmetry have
 no family quantum number. This assumption has the advantage of economy in the
 Higgs sector, preserves the gauge coupling unification, and avoids flavor-changing effects, but it requires the top quark
 mass to stem from a non-renormalizable coupling. 
 
 Second, each fermion field is a triplet ${\bf 3}$ under $\m Z_7 \rtimes \m Z_3$ and appears with
 its own antitriplet  familon scalar field $\overline{\bf 3}$. As a result, the Yukawa couplings
 are of the form 

$$\psi\psi'\,H\overline\xi\,\overline\xi' \ , $$
where $\psi$ denotes the fermions,  $\overline \xi$ the familons, and $H$ the Higgs fields. Guided by  $SU(5)$ unification, we associate a $\m Z_7 \rtimes \m Z_3$  antitriplet familon field to each of the fermion representations $\psi^{}_{\bf 10}$, $\psi^{}_{\bf \overline 5}$  and~$\psi^{}_{\bf 1}$,
as follows:

$$
\psi^{}_{\bf 10} ~ \rightarrow ~ \overline\varphi^{}_{\bf 10} \ ,\qquad 
\psi^{}_{\bf \overline 5} ~ \rightarrow ~ \overline\varphi^{}_{\bf \overline 5}  \ ,\qquad 
\psi^{}_{\bf 1} ~ \rightarrow ~ \overline\varphi^{}_{\bf 1} \ .
$$
Although labeled by the $SU(5)$ fermion representations to which they couple, the familon fields are themselves $SU(5)$ singlets. 

Third, we assume that the familon fields take on vacuum expectation values which extremize the values of the invariants in their potential. Hence we look for vacuum values aligned along the directions discussed in the previous section. We will include different signs for the entries, neglecting for the moment invariants capable of setting their phases.  The consequences of these assumptions can then be analyzed in the different sectors of the theory.

\subsection{Neutrinos}
In the $SU(5)$ picture, the charged leptons and the down-type quarks are
treated on equal footing. We therefore assume that the large MNSP mixing
predominantly originates from the neutrino sector which is special also in the
sense that it allows for the Seesaw mechanism
\cite{Minkowski:1977sc}. 
Taking $\nu_R^c$
($\psi^{}_{\bf 1}$) and $\nu_L$ ($\psi^{}_{\bf \overline 5}$) to be triplets
under $\m Z_7 \rtimes \m Z_3$, we 
can work out the most general structure of the Majorana and the Dirac
mass matrices in the coupling scheme $({\bf 3} \otimes {\bf 3}')^{}_{fermion} \otimes
({\bf \overline 3} \otimes {\bf \overline 3'})_{familon}$. Here the two ${\bf \overline 3}$s
represent the familon fields associated with their $SU(5)$ fermion
representations. Suppressing the Higgs field which in our approach does
not affect the flavor structure, we respectively have for the Majorana and the Dirac couplings

$$
\psi^{}_{\bf 1} \:  \psi^{}_{\bf 1} ~ \overline{\varphi}^{}_{\bf 1} ~ \overline{\varphi}^{}_{\bf 1} \ , \qquad 
\psi^{}_{\bf \overline 5} ~  \psi^{}_{\bf 1} ~ \overline{\varphi}^{}_{\bf \overline 5} ~ \overline{\varphi}^{}_{\bf 1} \ .
$$
Introducing the vacuum values for the familon fields 

$$
\langle \overline \varphi^{}_{\bf 1} \rangle ~\propto~ \pmatrix{a \cr b \cr c} \ , \qquad
\langle \overline\varphi^{}_{\bf \overline 5} \rangle ~\propto~ \pmatrix{a' \cr b' \cr c'} \ , 
$$
the general mass matrices obtained from the above $\m Z_7 \rtimes \m Z_3$
invariant coupling scheme are,  for the right-handed neutrinos, 

$$
\mathcal{M}_{\mathrm{Maj}} ~ \propto ~ \pmatrix{
a^2 & \alpha \, a  b &  \alpha \, a  c \cr
\alpha \, a  b & b^2 & \alpha \, b  c \cr
\alpha \, a  c &  \alpha \, b  c & c^2} \ ,
$$
and  the Dirac neutrino mass matrix is of the form

$$
\mathcal{M}_{\mathrm{Dir}} ~ \propto ~ \pmatrix{
a'a & \gamma_1\,a'b+\gamma_2\,b'a &  \beta_1\,a'c+\beta_2\,c'a \cr
\beta_1\,b'a+\beta_2\,a'b & b'b & \gamma_1\,b'c+\gamma_2c'b \cr
\gamma_1\,c'a+\gamma_2\,a'c & \beta_1\,c'b+\beta_2b'c & c' c}  \ .
$$
The parameters $\beta_{1,2}$ and $\gamma_{1,2}$ correspond to the four
independent invariants of type $I^{(4)}_4$; for the Majorana coupling there is
only one invariant of this type. Notice that we do not care about the overall
mass scales at this point, the only requirement we have is that the seesaw
formula \cite{Minkowski:1977sc} 

$$
\mathcal{M}_{\nu} ~ = ~ -\; \mathcal{M}_{\mathrm{Dir}} \cdot
\mathcal{M}^{-1}_{\mathrm{Maj}}\cdot \mathcal{M}^T_{\mathrm{Dir}} \ ,
$$
be applicable. In particular, this means that the Majorana mass matrix has to be invertible. Thus

$$
a,b,c ~ \neq ~ 0 \ , \qquad \alpha ~ \neq ~ -\,\frac{1}{2} \ , \qquad
\alpha ~ \neq ~ 1 \ .
$$
This suggests that among possible vacua which extremize the invariants, the
preferred familon $\overline\varphi_{\bf 1}$  aligns in the direction (up to signs) 

$$
\langle \overline \varphi^{}_{\bf  1} \rangle ~\propto~ 
\pmatrix{1 \cr 1  \cr 1} \ , 
$$
rather than in  the alternative directions $(0,0,1)$ or $(0,1,-1)$. None of the right-handed neutrinos are massless, although two have the same mass.

Assume the neutrino Dirac mass matrix is  family-symmetric, which can be
obtained by using only  the two invariants, $I^{(4)}_3$ and the symmetrized invariant of type $I^{(4)}_4$, that are symmetric in both $({\bf 3} \otimes {\bf 3}')^{}_{fermion}$ and $({\bf  \overline 3} \otimes {\bf \overline 3}')_{familon}$. This is tantamount to setting $\beta \equiv\beta_1 = \beta_2 = \gamma_1 = \gamma_2$. 

The resulting Seesaw mass matrix depends on two parameters, $\alpha$ which enters the right-handed neutrino mass matrix, and $\beta$ in the Dirac matrix, as well as on the vacuum alignment for $\overline \varphi^{}_{\bf\overline 5}$. 

We now assume a particular vacuum alignment, and fix $\alpha$ and $\beta$ so as to obtain a Seesaw matrix of the $\m M_{\m{TB}}$ type. Most alignments yield either degenerate eigenvalues, or masses which contradict experiment or else require enormously fine tuning. Only one alignment seems to produce  the desired result:

$$ 
\langle \overline \varphi^{}_{\bf\overline 5} \rangle ~\propto \pmatrix{-1 \cr \hfill 1  \cr \hfill 1} \ .
$$ 
The minus sign in the vacuum alignment does not affect the familon potential.
Two of the criteria for tri-bimaximal mixing are satisfied, and the third
fixes $\beta$ to an integer $-1$. Two ranges of values for $\alpha$,
$-1.15<\alpha< -0.95$ and $-0.22<\alpha< -0.15$, reproduce satisfactory
neutrino masses with the normal hierarchy.  For instance, with integer $\alpha
= -1$, the mass matrices become

$$
\mathcal{M}_{\mathrm{Maj}} ~\propto~ \pmatrix{
\hfill 1 & -1 & -1 \cr
-1 & \hfill 1 & -1 \cr
-1 & -1 & \hfill 1} \ , \qquad
\mathcal{M}_{\mathrm{Dir}} ~\propto~ \pmatrix{
-1 & \hfill 0 & \hfill 0 \cr
\hfill 0 &\hfill  1 & -2 \cr
\hfill 0 & -2 & \hfill 1} \ ,
$$
leading to the effective Majorana neutrino mass matrix

\bean
\mathcal{M}_{\nu}  &\propto &  -
\pmatrix{ -1 &\hfill  0 &\hfill  0 \cr\hfill  0 &\hfill  1 & -2 \cr\hfill  0 & -2 &\hfill  1} \cdot
\pmatrix{\hfill 0 & -\frac{1}{2} & -\frac{1}{2} \cr -\frac{1}{2} &0& -\frac{1}{2} \cr
-\frac{1}{2} & -\frac{1}{2} & \hfill 0} \cdot  
\pmatrix{ -1 & \hfill 0 &\hfill  0 \cr\hfill  0 &\hfill  1 & -2 \cr\hfill  0 & -2 & \hfill 1}  \\[2mm]
& \propto & \frac{1}{2} \, 
\pmatrix{\hfill 0&\hfill 1&\hfill 1 \cr\hfill  1&-4&\hfill 5 \cr\hfill \hfill  1&\hfill 5&-4} \ .
\eean
Comparing with $\mathcal{M}_{\mathcal{TB}}$, we see that this light neutrino mass matrix is diagonalized by the tri-bimaximal mixing matrix
$\mathcal{U}_{\mathcal{TB}}$. The eigenvalues are (normal) hierarchically ordered with 

$$
\mathcal{M}^{\mathrm{diag}}_{\nu}  ~\propto ~
\mathrm{diag}\,(\,1\,,-2\,,\,9\,) \ , 
$$
yielding the ratio of the atmospheric and the solar mass scales

$$
\frac{m^2_\mathrm{atm}}{m^2_\mathrm{sol}} ~=~ \frac{m_3^2 - m_1^2}{m_2^2-m_1^2}
~=~ \frac{81-1}{4-1} ~=~26.7 \ .
$$
This  compares favorably  with experiments, since a global three-generation fit gives the following $3\sigma$
allowed ranges \cite{Schwetz:2006dh} 

$$
\left.
\barr{lll}
m^2_{\mathrm{atm}} & \in & [\,1.9\,,\,3.2\,]\cdot10^{-3}\,\mathrm{eV}^2 \\[2mm]
m^2_{\mathrm{sol}} & \in & [\,7.1\,,\,8.9\,]\cdot10^{-5}\,\mathrm{eV}^2 
\earr
\right\} ~~\longrightarrow ~~ \frac{m^2_\mathrm{atm}}{m^2_\mathrm{sol}} ~\in~
[\,21.3\,,\,45.1\,] \ .
$$
It is striking that this solution yields the same value for $\alpha$ and $\beta$, which  suggests that these two $\m Z_7\rtimes \m Z_3$ invariants appear in the {\em same} combination in both the Majorana and the Dirac sectors, an indication perhaps of a higher symmetry.  Note that the masses are not degenerate, therefore this solution remains stable if one allows for small corrections in the parameters $\{\alpha,\beta_1,\beta_2,\gamma_1,\gamma_2\}$. We also investigated cases where we {\it{anti}}symmetrize with respect to $({\bf 3} \otimes {\bf 3}')_{fermion}$ and/or $({\bf \overline 3} \otimes {\bf \overline 3}')_{familon}$. However, none of them gave an equally convincing solution.


\subsection{Charged Leptons and Quarks}
The charged mass matrices $\mathcal{M}_u$, $\mathcal{M}_d$  ($\mathcal{M}_l$)
are generated from the couplings 

$$
\psi^{}_{\bf 10} ~\psi^{}_ {\bf 10} ~ \overline{\varphi}^{}_{\bf 10} ~ \overline{\varphi}^{}_{\bf 10} \ , \qquad
\psi^{}_{\bf 10}~\psi^{}_{\bf\overline 5}~\overline{\varphi}^{}_{\bf 10}~\overline{\varphi}^{}_{\bf \overline 5} \ , 
$$
respectively. The Dirac matrices have the general form

$$
\mathcal{M}_{} ~ \propto ~ \pmatrix{
x'x & \gamma\,x'y+\gamma'\,y'x &  \beta\,x'z+\beta'\,z'x \cr
\beta\,y'x+\beta'\,x'y & y'y & \gamma\,y'z+\gamma'z'y \cr
\gamma\,z'x+\gamma'\,x'z & \beta\,z'y+\beta'y'z & z' z}  \ .
$$
Assuming that the $\m Z_7 \rtimes \m Z_3$ invariant $I^{(4)}_4$
is absent or strongly suppressed ($\beta,\beta',\gamma,\gamma'\ll 1$), these reduce to diagonal form 

$$
\mathcal{M}_u 
~\propto~\mathrm{diag}\,(\,{a''}^2 ,\,{b''}^2 ,\,{c''}^2 \,) \ , \qquad 
\mathcal{M}_d \,,\, \mathcal{M}_l 
~\propto~\mathrm{diag}\,(\,-{a''} ,\,{b''} ,\,{c''} \,) \ ,
$$
where we have used the value of $\langle\overline\varphi_{\bf\overline 5}\rangle$ of the previous section, and set

$$
\langle\overline \varphi^{}_{\bf 10} \rangle ~\propto~ 
\pmatrix{a'' \cr b'' \cr c''}  \ ,
$$
as the third vacuum alignment. 

We note that  the quadratic dependence on the familon vacuum expectation values generates a  hierarchy between $m_t$ and $m_c$ that is automatically much larger than that between $m_b$ and $m_s$. This is in quantitative agreement since the ratios of the fermionic masses at the GUT scale are given in terms of the Wolfenstein parameter $\lambda_c$ by 
\cite{Binetruy:1994ru,
Nir:1995bu}
\bean
m_u~:~m_c~:~m_t & \sim & \lambda_c^8 ~:~ \lambda_c^4 ~:~ 1 \ , \\
m_d~:~m_s~:~m_b & \sim & \lambda_c^4 ~:~ \lambda_c^2 ~:~ 1 \ , \\
m_e~:~m_\mu~:~m_\tau & \sim & \lambda_c^{4\:\mathrm{or}\:5} ~:~ \lambda_c^2 ~:~ 1 \ .
\eean
The dominant masses of the top quark, bottom quark and the tau lepton are clearly reproduced by the allowed vacuum alignment of the familon field $\overline \varphi^{}_{\bf 10}$ 

$$
\langle\overline \varphi^{}_{\bf 10} \rangle ~\propto~ 
 \pmatrix{0 \cr 0 \cr 1} \ ,
$$
but this would also make the two charged leptons massless. Because of this degeneracy, the tri-bimaximal matrix is no longer uniquely determined. One possible solution to this problem is to assume that the vacuum alignment of the familon fields is altered by Cabibbo effects; for example, one could imagine a new alignment 

$$
\langle\overline \varphi^{}_{\bf 10} \rangle ~\propto~ 
 \pmatrix{\lambda_c^4 \cr \lambda_c^2 \cr 1} \ ,
$$
which leads to a satisfactory description of the charged fermion masses of the second and first generations. 

This alteration of $\langle\overline \varphi^{}_{\bf 10} \rangle$  does not affect the MNSP matrix, although one can in principle expect similar corrections to $\langle\overline \varphi^{}_{\bf \overline 5} \rangle$ and $\langle\overline \varphi^{}_{\bf 1} \rangle$. To conclude, our model generates an approximation with tri-bimaximal MNSP mixing in the neutrino sector and no quark CKM mixing.

Corrections to this approximate model would have to include the Cabibbo size mixing in the CKM matrix. This might be achieved by slightly switching on
the invariants of type $I^{(4)}_4$. Doing so in the down quark sector, and thus automatically also in the charged lepton sector, one would get deviations from
tri-bimaximal mixing in the MNSP matrix as well.  This is clearly important in determining the size of the CHOOZ angle. 

Although it is beyond the scope of this work to list the many ways in which this can be achieved, we illustrate our point with one example. Consider the case where the mixing of the charged leptons is of the same structure as $\m
U_{CKM}$, 

$$
\m U_{-1} ~\approx ~ 
\pmatrix{1 & \lambda_e &0 \cr 
-\lambda_e & 1 & 0 \cr
0 & 0 & 1 }~+~\m O(\lambda_e^2) \ ,
$$
where 

$$
\lambda^{}_e\approx~\sqrt{\frac{m_e}{m_\mu}}\ ,$$
is the leptonic equivalent \cite{Harvey:1980je} 
of the Gatto relation. Then we find to linear order in $\lambda_e$
$$
\m U_{MNSP} ~\approx ~ \m U^\dagger_{-1} \, \m U^{}_{\m{TB}} ~\approx ~
\pmatrix{
\sqrt{\frac{2}{3}} \left( 1+ \frac{\lambda_e}{2}\right)   &
   \frac{1}{\sqrt{3}} \left( 1 - \lambda_e \right) &  
    \frac{1}{\sqrt{2}} \lambda_e \cr 
-\,\frac{1}{\sqrt{6}} \left( 1- 2 \lambda_e \right) & 
   \frac{1}{\sqrt{3}} \left( 1 + \lambda_e \right) & 
   - \, \frac{1}{\sqrt{2}}  \cr 
-\,\frac{1}{\sqrt{6}} & 
 \frac{1}{\sqrt{3}}  & 
    \frac{1}{\sqrt{2}} } \ .
$$
In this case, the $(1,3)$-entry would correspond to an angle
$\theta_{13} \approx 2.8^\circ$, which is too small to be testable by the
forthcoming Double-CHOOZ experiment \cite{Ardellier:2006mn}. It will also
change the solar neutrino mixing from its tri-bimaximal value of $35.3^\circ$
to $32.4^\circ$, which is to be compared with the $1\sigma$ range of the global
three-generation fit \cite{Schwetz:2006dh}: $\theta_{12} \in [31.3^\circ
\,,\,34.4^\circ ]$.

Our approximation  attributes a different origin to the mixing parameters and
the quark mass ratios. If, as the texture zeros
\cite{Wilczek:1977uh} 
and the
Froggatt-Nielsen \cite{Froggatt:1978nt} approaches suggest, the Cabibbo angle generates {\em both} quark mixings {\em and} the charged fermion masses of the two lightest families, it must be treated in the context of degenerate perturbation theory, where the mixing is determined by the form of the perturbation.  
 

\section{Summary}
The quest for a convincing explanation of the triplication of chiral families
is still ongoing. Many authors have constructed models of flavor adopting
Non-Abelian finite groups of their choice. Although these 
 all give rise to tri-bimaximal mixing in the neutrino sector, it
remains unclear why some particular group should be preferred to others.

In this letter, we have argued for focusing on the unique {\it simple} finite
subgroup of $SU(3)$ with a complex triplet representation, $\m P \m S \m
L_2(7)$. In particular, we have drawn attention to one of its  subgroups,  the
Frobenius group $\m Z_7 \rtimes \m Z_3$ as a possible family group. Using only
the basic alignment vectors which extremize its invariants, one readily
obtains tri-bimaximal mixing in the neutrino sector. 

A particularly suggestive model, in which the Majorana as well as the Dirac
mass matrices are derived from the {\em same}  sum of two independent group
invariants with  {\it integer} relative coefficients ($\alpha=\beta=-1$),  yields  phenomenologically viable neutrino mass ratios.  It remains to be seen whether this can be understood in terms of a higher symmetry. Quark and  charged lepton masses for the first and the second
families can be generated by corrections to the vacuum alignments,  without necessarily inducing CKM mixing. 

Expected deviations from tri-bimaximal mixing ("Cabibbo haze") will then determine the size of the CHOOZ angle, and shift the tri-bimaximal values of the atmospheric and solar angles. A simple $SO(10)$-inspired scenario of this type is presented, in which the CHOOZ angle is very small, and  the solar angle is shifted towards its experimental  value.

\section*{Acknowledgments}
We are grateful for inspiring discussions with G. G. Ross.
The work of one of the authors (C.L.) is supported by the University of
Florida through the Institute for Fundamental Theory and that of two of the
authors (S.N. and P.R.) is supported by the Department of Energy Grant
No. DE-FG02-97ER41029.


\end{document}